\documentclass{llncs}
\pdfoutput=1
\usepackage{amsfonts}
\usepackage{times}
\usepackage{graphicx}
\graphicspath{{images/}}
\usepackage{url}
\urldef{\mailsa}\path|{galeas, freisleb}@informatik.uni-marburg.de|
\urldef{\mailsb}\path|ralph.kretschmer@raytion.com| 

\newcommand{\beq}{\begin{equation}}
\newcommand{\eeq}{\end{equation}}
\newcommand{\md}{{\rm d}}
\newcommand{\const}{\rm const}
\newcommand{\mcal}[1]{\mathcal{#1}}
\newcommand{\mrm}[1]{\mathrm{#1}}
\newcommand{\simil}{\mathop{\mathrm{sim}}}

\begin{document}

\title{Information Retrieval via Truncated Hilbert-Space Expansions}

\author{Patricio Galeas\inst{1} \and Ralph Kretschmer\inst{2} \and Bernd Freisleben\inst{1}}
 
\institute{Dept. of Mathematics and Computer Science, University of Marburg, \\
Hans-Meerwein-Str. 3, D-35032 Marburg, Germany\\
\mailsa
\and
Raytion GmbH, \\
Kaiser-Friedrich-Ring 74, D-40547 D\"{u}sseldorf, Germany\\
\mailsb}

\maketitle
\thispagestyle{empty}

\begin{abstract}
In addition to the frequency of terms in a document collection, the distribution
of terms plays an important role in determining the relevance of documents.
In this paper, a new approach for representing
term positions in documents is presented. The approach allows
an efficient evaluation of term-positional information at query evaluation time.
Three applications are investigated: a function-based
ranking optimization representing a user-defined document region, a
query expansion technique based on overlapping the term distributions in the
top-ranked documents, and cluster analysis of terms in documents. Experimental results demonstrate the effectiveness of the
proposed approach.

\end{abstract}


\section{Introduction}

The information retrieval (IR) process has two main stages. The first stage is the
indexing stage in which the documents of a collection are processed
to generate a database (index) containing the information about the terms of all
documents in the collection. The index generally stores only term frequency
information, but in some cases positional information of terms is also
included, substantially increasing the memory requirements of the system.

In the second stage of the IR process (query evaluation), the user sends a
query to the system, and the system responds with a ranked list of relevant
documents. The implemented retrieval model determines how the relevant documents
are calculated.
Standard IR models (e.g.\ TFIDF, BM25) use the frequency of terms as the main
document relevance criterion, producing adequate quality in the ranking and
query processing time. Other approaches, such as proximity queries or passage
retrieval, complement the document relevance evaluation using term
positional information. This additional process, normally performed at query
time, generally improves the quality of the results but also slows down the
response time of the system.
Since the response time is a critical issue for the acceptance of an
IR system by its users, the use of time-consuming algorithms to evaluate term-positional
information at query time is generally inappropriate.

The IR model proposed in this paper shifts the complexity of processing the
positional data to the indexing phase, using an abstract representation of the
term positions and implementing a simple mathematical tool to
operate with this compressed representation at query evaluation time. Thus,
although query processing remains simple, the use of term-positional information
provides new ways to optimize the IR process. Three applications are investigated: a function-based
ranking optimization representing a user-defined document region, a
query-expansion technique based on overlapping the term distributions in the
top-ranked documents, and cluster analysis of terms in documents. Experimental results demonstrate the effectiveness of the
proposed approach for optimizing the retrieval
process.

The paper is organized as follows. Section \ref{relwork} discusses related work. 
Section \ref{approach} presents the proposed approach for representing term positions based on truncated Hilbert space
expansions. In Section \ref{applying_model}, applications of the approach are described. 
Section \ref{conc} concludes the paper and outlines areas for future work.

\section{Related Work}
\label{relwork}
An early approach to apply term-positional data in IR is the work of Attar and
Fraenkel \cite{Attar77}. The authors propose different models to generate clusters of
terms related to a query (searchonyms) and use these clusters in a local
feedback process. In their experiments they confirm that metrical methods based
on functions of the distance between terms are superior to methods based merely
on weighted co-occurrences of terms. There are several other approaches that use metrical
information \cite{Beigbeder05,Tao07}.

One of the first approaches using abstract representations of term distributions
in documents is Fourier Domain Scoring (FDS), proposed by Park et al.\
\cite{Park04}. FDS performs a separate magnitude and phase analysis of term
position signals to produce an optimized ranking. It creates an index based on
page segmentation, storing term frequency and approximated positions in the
document. FDS processes the indexed data using the \textit{Discrete Fourier
Transform} to perform  the corresponding spectral analysis.

A recent approach based on an abstract representation of term position is
Fourier Vector Scoring (FVS) \cite{Galeas09}. It represents the term
information (Fourier coefficients) directly as an $n$-dimensional vector using
the analytic Fourier transform, permitting an immediate and simple term
comparison process. 

\section{Analyzing Term Positions}
\label{approach}
In this section, a general mathematical model
to analyze term positions in documents is presented, making it possible to effectively use
the term-positional information at query evaluation time.

Consider a document $D$ of length $L$ and a term $t$ that appears in $D$.
The distribution of the term $t$ within the document is given by
the set $\mcal{P}_t$ that contains all positions of $t$, where all terms are
enumerated starting with 1 for the first term and so on. For example,
a set $\mcal{P}_t = \{2, 6\}$ represents a tern that is located at the second
and sixth position of the document body. A characteristic function
\beq\label{eq5}
f^{(t)}(x) = \left\{ \begin{array}{ll}
1 & \mbox{for } x \in [p - 1, p] \mbox{ if } p \in \mcal{P}_t \\
0 & \mbox{otherwise}
\end{array} \right. ,
\eeq
defined for $x \in [0, L]$, is assigned to $\mcal{P}_t$.

The proposed method consists of approximating this characteristic function by
an expansion in terms of certain sets of functions.  In order to do so, some
concepts of functional analysis are introduced.
Details can be found in the book of Yosida \cite{Yosida80}.

\subsection{Expansions in Hilbert Spaces}

A Hilbert space $\mcal{H}$ is a (possibly infinite-dimensional) vector space
that is equipped with a scalar product $\langle .,. \rangle$, i.~e.\ two
elements $f, g \in \mathcal{H}$ are mapped to a real or complex number
$\langle f, g \rangle$. We only consider real scalar products here.

An example of a Hilbert space is the  space $L_2([0, L])$ defined as the
set of all functions $f$ that are {\em square-integrable\/} in the interval
$[0, L]$, i.~e.\ functions for which
$\int_0^L (f(x))^2 \, \md x < \infty \,.$
In this vector space, the addition of two functions $f$ and $g$, and the
multiplication of a function $f$ by a scalar $\alpha \in \mathbb{R}$ are defined
point-wise: $(f + g)(x) = f(x) + g(x) \,,\ (\alpha f)(x) = \alpha f(x) \,$.
The scalar product in $L_2([0, L])$ is defined by
\beq\label{eq2}
\langle f, g \rangle = \int\limits_0^L f(x) g(x) \, \md x \,.
\eeq
Two vectors with vanishing scalar product are called {\em orthogonal}.

The scalar product induces a {\em norm\/} (an abstract measure of length)
\beq
\|f\| = \sqrt{\langle f, f \rangle} \geq 0 \,.
\eeq
With the help of this norm, the notion of {\em convergence\/} in $\mathcal{H}$
can be defined: A sequence $f_0, f_1, \ldots$ of vectors of $\mcal{H}$ is said
to converge to a vector $f$, symbolically $\lim_{n \to \infty} f_n = f$, if
$\lim_{n \to \infty} \|f_n - f\| = 0$. This allows to define an expansion of a
vector $f$ in terms of a set of vectors
$\{\varphi_0, \varphi_1, \allowbreak \ldots\}$. One writes
\beq
f = \sum_{k = 0}^\infty \gamma_k \varphi_k \,,
\eeq
where the $\gamma_k$ are real numbers, if the sequence
$f_n = \sum_{k = 0}^n \gamma_k \varphi_k$ of finite sums converges to $f$. This
kind of convergence is called {\em norm convergence}.

Of particular importance are so-called complete, orthonormal sets
$\{\varphi_0, \varphi_1, \ldots\}$ of functions in $\mathcal{H}$. They have the
following properties: (a)~The $\varphi_i$ are mutually orthogonal and normalized
to unity:
\beq
\langle \varphi_n, \varphi_m \rangle = \delta_{n m}
= \left\{ \begin{array}{ll}
1 & \mbox{for } n = m \\
0 & \mbox{for } n \neq m
\end{array} \right. \,
\eeq
(b)~The $\varphi_i$ are {\em complete\/}, which means that every vector of
the Hilbert space can be expanded into a convergent sum of them.

Important properties of expansions in terms of complete orthonormal sets are:
(a)~The expansion coefficients $\gamma_k$ are given by
\beq\label{eq8}
\gamma_k = \langle \varphi_k, f \rangle \,.
\eeq
(b)~They fulfill
\beq\label{eq4}
\sum_{k = 0}^n \gamma_k^2 \leq \|f\|^2 \mbox{ for all } n
, \mbox{ and }
\sum_{k = 0}^\infty \gamma_k^2 = \|f\|^2
\eeq
(Bessel's inequality and Parseval's equation).

Given two expansions $f = \sum_{k = 0}^\infty \gamma_k \varphi_k$,
$g = \sum_{k = 0}^\infty \gamma_k' \varphi_k$, the scalar product can be
expressed as
\beq
\langle f, g \rangle = \sum_{k = 0}^\infty \gamma_k \gamma_k' \,.
\eeq
If the expansion coefficients are combined into coefficient vectors
$\vec{c} = (\gamma_0, \gamma_1, \ldots)$,
$\vec{c}' = (\gamma_0', \gamma_1', \ldots)$, the preceding equation takes the
form
$\langle f, g \rangle = \vec{c} \cdot \vec{c}'$.

The Fourier expansions considered by Galeas et al.\ \cite{Galeas09} are an example of such an
expansion. The functions
\begin{equation}
\varphi^\mrm{Fo}_0(x)  =  \frac{1}{\sqrt{L}} \,,\
\varphi^\mrm{Fo}_{2 k - 1}(x)
= \sqrt{\frac{2}{L}} \sin \left( \frac{2 \pi k}{L} \right) \,,\
\varphi^\mrm{Fo}_{2 k}(x)
 =  \sqrt{\frac{2}{L}} \cos \left( \frac{2 \pi k}{L} \right)
\label{eq10}
\end{equation}
($k > 0$) form a complete orthonormal set in $L_2([0, L])$, leading to an
expansion
\beq
f(x) = \frac{a_0}{\sqrt{L}} + \sqrt \frac{2}{L} \sum_{k = 1}^\infty
\left[ a_k \cos \left(\frac{2 \pi k x}{L} \right)
+ b_k \sin \left( \frac{2 \pi k x}{L} \right) \right] \,,
\eeq
where $a_0=\gamma_0$ and $a_k = \gamma_{2 k}$, $b_k = \gamma_{2 k - 1}$ for 
$k>0$.

\begin{figure}[t]
\begin{center}
\begin{tabular}{cc}
\includegraphics[width=0.5\linewidth]{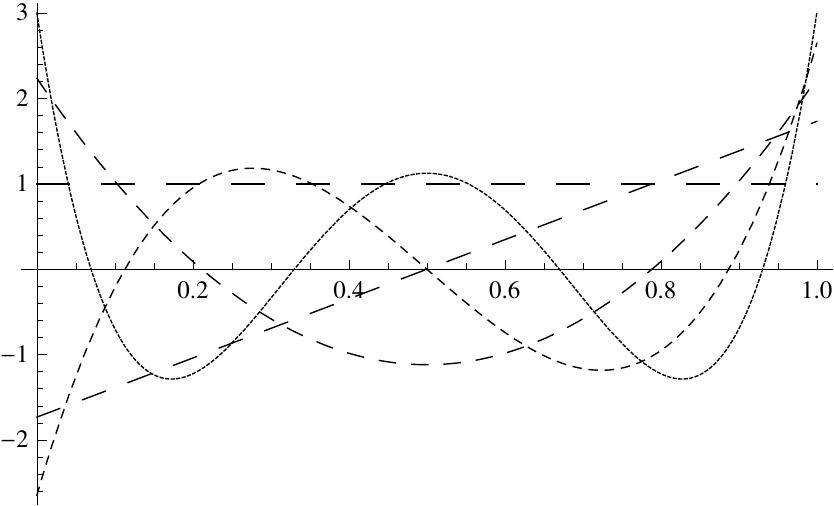} &
\includegraphics[width=0.5\linewidth]{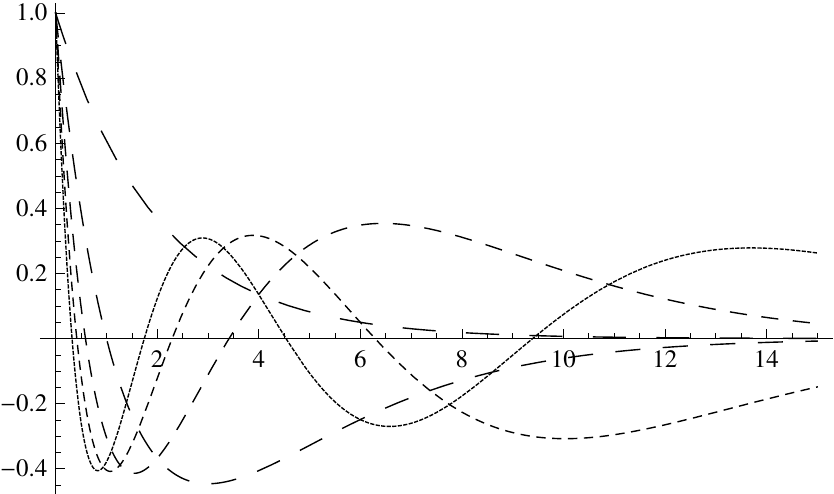}
\end{tabular}
\end{center}
\caption{\textit{Left}: Shifted Legendre polynomials  $\varphi^\mrm{Le}_k(x)$ for
$0\le k \le 4$. \textit{Right}: The expansion functions (\ref{eq17}) for $0 \le k \le 4$
and $\lambda=1$}
\label{legendre-laguerre}
\end{figure}

Another complete set of orthonormal functions of $L_2([0, L])$ is given by
\beq
\label{eq16}
\varphi^\mrm{Le}_k(x) = \sqrt{\frac{2 k + 1}{L}} P^*_k(x / L) \,,\ k \geq 0 \,,
\eeq
where the $P^*_k(x)$ are so-called {\em shifted Legendre polynomials\/}
\cite{Abramowitz}. These polynomials are of order $k$. The first few of them are
$P^*_0(x) = 1$, $P^*_1(x) = 2 x - 1$,
$P^*_2(x) = 6 x^2 - 6 x + 1$,
$P^*_3(x) = 20 x^3 - 30 x^2 + 12 x - 1$.
Fig.~\ref{legendre-laguerre} (left) shows $\varphi^\mrm{Le}_k(x)$ for
$0\le k \le 4$ in the range $x \in [0, L]$ for $L=1$.

Another example that will be used later is a complete set for the space
$L_2(\mathbb{R}_+)$ (the space of square-integrable functions for
$0 \leq x < \infty$):
\beq
\label{eq17}
\varphi^\mrm{La}_k(x)
= \frac{e^{- x / (2 \lambda)}}{\sqrt{\lambda}} L_k(x / \lambda)
\,,\ k \geq 0 \,.
\eeq
Here, $\lambda$ is a positive scale parameter and the $L_k(x)$ are {\em Laguerre
polynomials\/} \cite{Abramowitz}, the first few of which are
$L_0(x) = 1$,
$L_1(x) = - x + 1$,
$L_2(x) = x^2/2 - 2 x + 1$,
$L_3(x) = - x^3/6 + 3x^2/2 - 3 x + 1$,
see Fig.~\ref{legendre-laguerre} (right).

\subsection{Truncated Expansions of Term Distributions}
\label{truncation}

As explained above, the finite sums $f_n = \sum_{k = 0}^n \gamma_k \varphi_k$
converge to the function $f$ in the sense of norm convergence. As a consequence
of Bessel's inequality (\ref{eq4}) they approximate $f$ increasingly better for
increasing $n$. An essential ingredient for the following discussion is to
consider a truncated expansion, i.~e.\ the mapping
\beq
P_n: f^{(t)} \mapsto f_n^{(t)} \,,
\eeq
which associates to a term distribution $f^{(t)}$ of the form (\ref{eq5}) its
finite-order approximation $f_n^{(t)}$ in terms of some complete orthonormal set
for some order $n$.
\begin{figure}[t]
\centering
\includegraphics[width=0.5\linewidth]{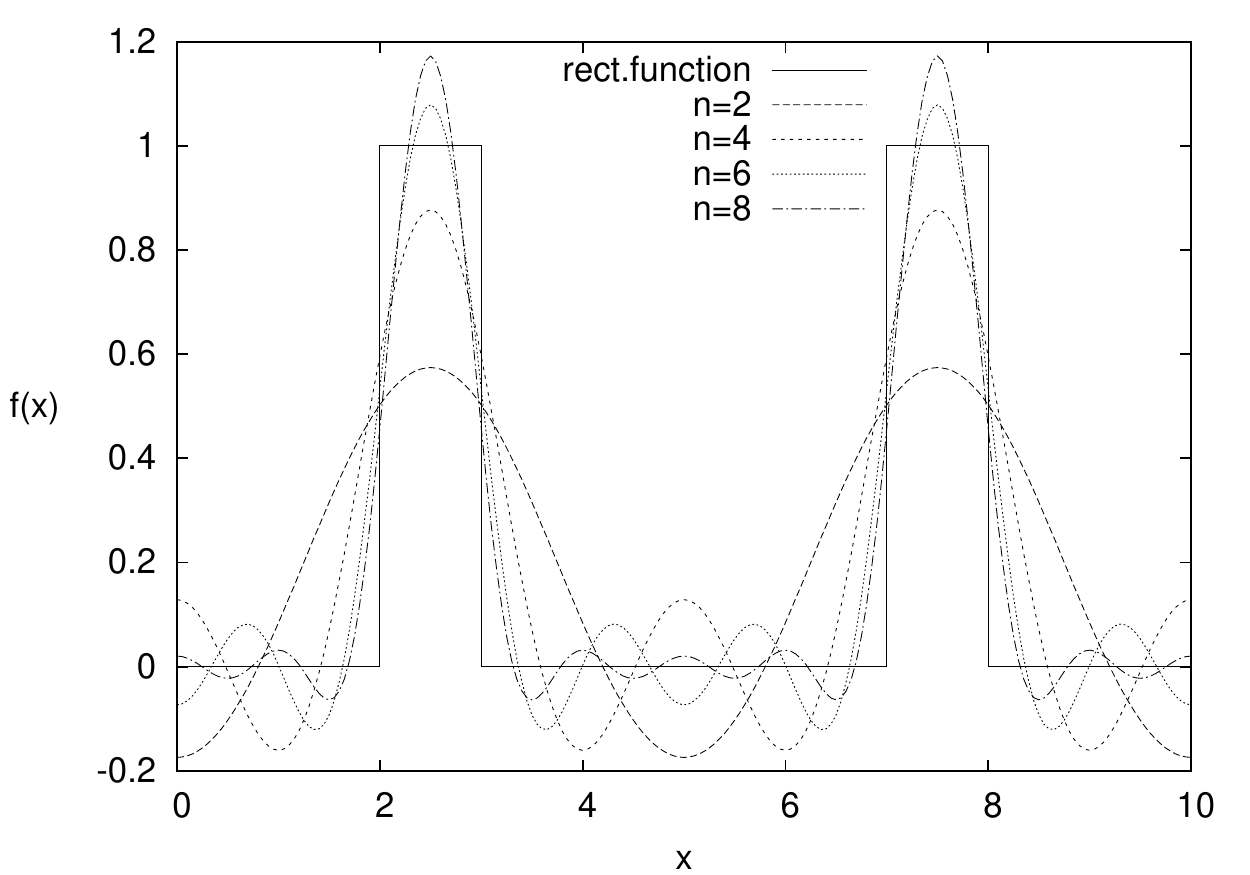}
\caption{Fourier distribution of $\mathcal{P}_t=\{3, 8\}$ in document $D$, using
different Fourier orders $n$.}
\label{word_positions}
\end{figure}

Figure \ref{word_positions} shows an example for the Fourier expansion. One can
observe the characteristic broadening effect generated by the reduction of the
expansion order (truncation).

The $L_2$ scalar product of two truncated term distributions $f_n$ and $g_n$,
\beq
\langle f_n, g_n \rangle = \int f_n(x) g_n(x) \, \md x
\eeq
has the meaning of an {\em overlap integral}: The integrand is only large in
regions in which both functions $f_n(x)$ and $g_n(x)$ are large, so that
$\langle f_n, g_n \rangle$ measures how well both functions overlap in the whole
integration range.

Given $f_n$ and $g_n$, two truncated term distributions describing the term
positions and their neighborhood in a certain document, we introduce the concept
of {\em semantic interaction range}: Two terms that are close to each other
present a stronger interaction because their truncated distributions have a
considerable overlap.
This semantic interaction range motivates the following definition of the
{\em similarity\/} of two term distributions $f$ and $g$: For some fixed order
$n$,  one sets
\beq\label{eq6}
\simil(f, g) = \langle f_n, g_n \rangle = \langle P_n f, P_n g \rangle \,.
\eeq
In this definition, the truncation $P_n: f \mapsto f_n$ is essential, because
the  original term distributions $f$ and $g$ are always orthogonal if they
describe  two different terms. This is so because different terms are always at
different  positions within a document, so that their overlap always vanishes.

Definition (\ref{eq6}) is only one possibility. In fact, any definition
based on the scalar product $\langle f_n, g_n \rangle$ can be utilized. For
example, in Galeas et al.\ \cite{Galeas09} a cosine definition
$\cos \vartheta = \frac{\langle f_n, g_n \rangle}{\|f_n\| \|g_n\|}$
has been used. Another choice is the norm difference
\begin{equation}
\|f_n - g_n\|  =  \left( \int (f_n(x) - g_n(x))^2 \,\md x \right)^{1 / 2}
 = \sqrt{\|f_n\|^2 + \|g_n\|^2 - 2 \langle f_n, g_n \rangle} \,.
\label{eq12}
\end{equation}
Using different measures based on
$\langle f_n, g_n \rangle$, we have found no significant differences in the
final retrieval results in several experiments.

The scalar product of the truncated distributions can be easily calculated using
the coefficient vectors: If the original distributions $f$ and $g$ have the infinite-dimensional coefficient vectors
$\vec{c} = (\gamma_0, \gamma_1, \ldots)$ and
$\vec{c}' = (\gamma_0', \gamma_1', \ldots)$, respectively, then the truncated
distributions $f_n$ and $g_n$ have the $(n + 1)$-dimensional coefficient vectors
$\vec{c}_n = (\gamma_0, \gamma_1, \ldots, \gamma_n)$ and
$\vec{c}'_n = (\gamma_0', \gamma_1', \ldots, \gamma_n)$, resp., and their
scalar product is the finite sum
\beq\label{eq15}
\langle f_n, g_n \rangle = \vec{c}_n \cdot \vec{c}_n'
= \sum_{k = 0}^n \gamma_k \gamma_k' \,.
\eeq

\subsection{The Semantic Interaction Range}

In this section, a precise definition of the semantic interaction range is given.

In abstract terms, the truncation $P_n: f \mapsto f_n$ is a filtering or a
{\em projection}: In the expansion
$f(x) = \sum_{k = 0}^\infty \gamma_k \varphi_k(x)$ the components $\varphi_k$
for $k > n$ are filtered out, which amounts to a projection of $f$ onto the
components $\varphi_0, \ldots, \varphi_n$. Thus, $P_n$ is a projection
operator in the Hilbert space.
To derive a closed expression for the operator $P_n$, one combines
$(P_n f)(x) = f_n(x) = \sum_{k = 0}^n \gamma_k \varphi_k(x)$,
with (\ref{eq8}) to obtain
\begin{equation}
(P_n f)(x)
 =  \sum_{k = 0}^n \left( \int \varphi_k(y) f(y) \,\md y \right) \varphi_k(x)
= \int \left( \sum_{k = 0}^n \varphi_k(y) \varphi_k(x) \right) f(y) \,\md y\,.
\label{eq9}
\end{equation}
One can write the last expression as $\int p_n(y, x) f(y) \,\md y$
with the {\em projection kernel\/}
\beq
p_n(y, x) = \sum_{k = 0}^n \varphi_k(y) \varphi_k(x)
\eeq
as an integral representation of $P_n$ in the sense of a
convolution. It has the advantage that one can study the properties of the
truncation independently of the function $f$.

The width of $p_n(y, x)$ as a function of $x$ is a lower bound for the width of
a truncated expansion of a term located at $y$. Therefore, this width will be
used as the semantic interaction range for a term at position $y$.

\begin{figure}[t]
\begin{center}
\begin{tabular}{cc}
\includegraphics[width=0.5\linewidth]{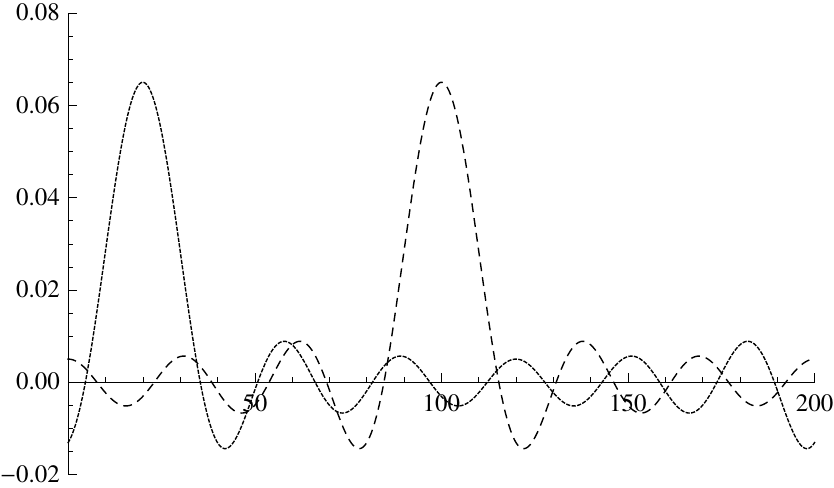} &
\includegraphics[width=0.5\linewidth]{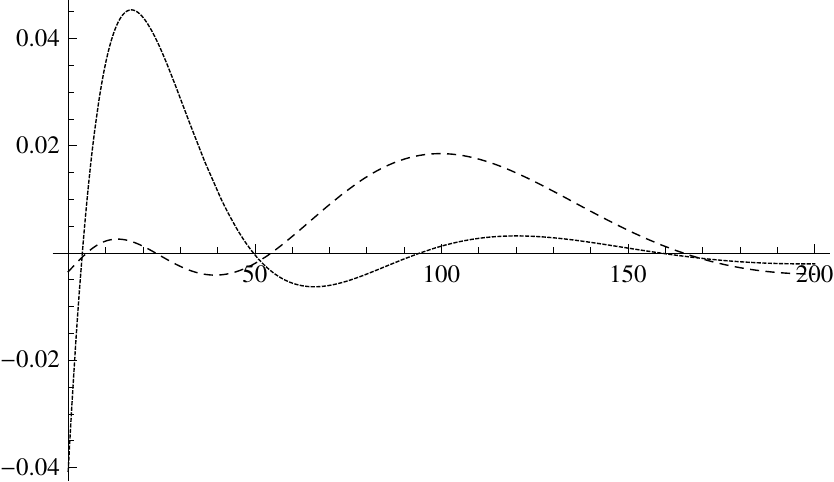}
\end{tabular}
\end{center}
\caption{\textit{Left}: Projection kernel for the Fourier expansion showing the semantic
interaction range for two terms at the positions 20 and 100, for $n=6$ and
$L=200$. \textit{Right}: Projection kernel for the expansion in terms of Laguerre
polynomials showing the semantic interaction range for two terms at the
positions 20 and 100, for $n=6$ and $\lambda=15$.
\label{fourier-laguerre-projectors}}
\end{figure}

For the Fourier expansion, $p_{2 k}$ is given by
\begin{equation}
p^\mrm{Fo}_{2 k}(y, x) = \frac{\cos(4 \pi k (y - x) / L)
- \cos(2 \pi (2 k + 1) (y - x) / L)}{L (1 - \cos(2 \pi (y - x) / L))} \,.
\end{equation}
(We consider only even orders $n = 2 k$, because for these orders the
expansion consists of an equal number of sine and cosine terms, see
(\ref{eq10}).) The maximum of $p^\mrm{Fo}_{2 k}(y, x)$ is at $x = y$ and
the two zeros closest to the maximum are at $x = y \pm L / (2 n + 1)$.
Thus, the semantic interaction range for a Fourier expansion of order $n$ may be
defined to be
\beq
\varrho^\mrm{Fo}_n = \frac{2 L}{2 n + 1} \,.
\eeq
Fig.~\ref{fourier-laguerre-projectors} (left) shows $p^\mrm{Fo}_6(20, x)$ and
$p^\mrm{Fo}_6(100, x)$ for $L = 200$.

For the expansions in terms of Legendre and Laguerre polynomials, the
projection kernels can be calculated with the Christoffel-Darboux equation
\cite{Abramowitz}. The results are
\beq
p^{i}_n(y, x) = \alpha^i_n
\frac{\varphi^i_{n + 1}(y) \varphi^i_n(x) - \varphi^i_n(y) \varphi^i_{n + 1}(x)}
{y - x} \,,
\eeq
$i =$ Le, La, with $\alpha^\mrm{Le}_n = (L / 2) (n + 1) / (2 n + 1)$
and $\alpha^\mrm{La}_n = - \lambda (n + 1)$. These kernels are no longer
functions of $y - x$, meaning that the broadening of a term
distribution depends on the position $y$ of the term distribution within the
document.

Fig.~\ref{fourier-laguerre-projectors} (right) shows the projection kernel
$p^\mrm{La}_6(y, x)$ for $y = 20$ and $y = 100$. One can see that the spatial
resolution of the truncated expansion decreases for terms that are far away from
the beginning of the document.

\section{Applications}
\label{applying_model}
The goal of our approach is to shift the complexity of processing the positional
data from the query evaluation phase to the (not time critical) indexing phase,
reducing the ranking optimization via term positions to a simple mathematical
operation.

Hence, we propose to calculate the expansion coefficients $\gamma_k$ of
the term distributions in the indexing phase and to store this abstract term
positional information in the index.
This permits a considerably faster query evaluation, compared with methods that
use the raw term-positional information.

Thus, the index contains an $(n + 1)$-dimensional coefficient vector
$\vec{c}_n = (\gamma_0, \gamma_1, \ldots,\allowbreak \gamma_n)$ for each term and
each document in the collection. The $\gamma_k$ are calculated analytically
via (\ref{eq8}).
To give an example of the complexity involved,
\begin{equation}
\gamma_k = \sum_{p \in \mcal{P}_t} \sum_{j = 0}^k \alpha_j
\left[ \left( \frac{p}{L} \right)^{j + 1}
- \left( \frac{p - 1}{L} \right)^{j + 1} \right]
\end{equation}
with $\alpha_j = \sqrt{(2 k + 1) L} \, a_j / (j+1)$ is the
expression for the expansion coefficients in the case of the expansion in terms
of Legendre polynomials, cf.\ (\ref{eq5}). (The $a_j$ are the polynomial
coefficients of the shifted Legendre polynomial of order $k$.) Calculations of
this kind can be easily performed in the indexing stage.

The retrieval scenarios that we have investigated are:
(a) ranking optimization based on {\em user-defined objective functions}
and (b) {\em query expansion based on term-positional
information} \cite{Galeas09}, and (c) cluster analysis of terms in documents. They all involve a calculation
of the similarity of term distributions.

\subsection{Ranking Optimization}
\begin{figure}[t]
\centering
\includegraphics[width=0.80\linewidth]{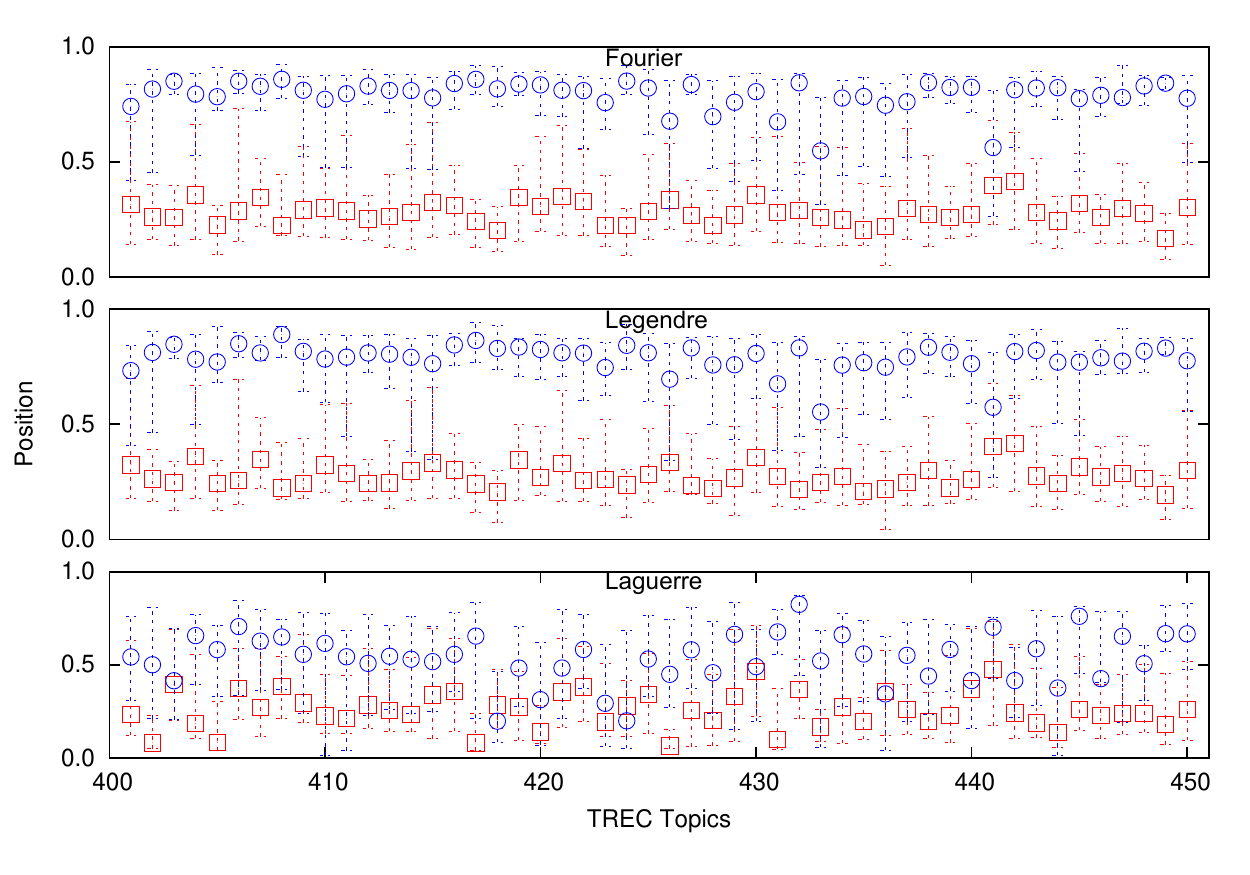}
\caption{Objective function performance for the Fourier, Legendre and Laguerre
models. The $x$ axis shows the TREC topics 400 to 450, the $y$ axis is the term
position relative to the normalized document length. The circle and the rectangle
bounds represent the range of the query term positions for the objective functions
$1|3$ and $3|3$ respectively.}
\label{goal-functions}
\end{figure}

The first scenario states document ranking as an optimization problem
that is based on the query term distribution function $f_{q,d}$ and a
user-defined objective function $f_\mathrm{o}$ representing the optimal
query term distribution in the document body:
\begin{equation}
\label{objective_ranking_function}
Maximize\left\{ \simil(f_{q,d},f_\mathrm{o}) \right\} \hspace{1cm}
\forall f_{q,d} \in A
\end{equation}
where $A$ represents the query term distributions in a document set,
$f_{q,d}$ is the query term distribution function for query $q$ in
document $d$, and $f_\mathrm{o}$ is a user-defined objective function,
representing the optimal query term distributions for the documents in
the document ranking.
Experiments based on the TREC-8 collection and the software Terrier \cite{Ounis06}, 
carried out to order $n=6$, show the accuracy of the term distributions in a 
ranking based on user-defined objective functions.
As depicted in Figure \ref{goal-functions}, the Fourier and Legendre
models present a high accuracy for the distribution of query terms in the top-20
ranked documents, based on two different objective functions: The first function (denoted $f_\mrm{o}=1|3$) selects terms located in the first third of the document,
and the second ($f_\mrm{o}=3|3$) selects terms located in the last third of the
document \cite{Galeas09}.

\subsection{Query Expansion}
The second scenario considers the \textit{top-r} documents
$D=\left\lbrace d_1, d_2,\dots,d_r\right\rbrace$ of an initial ranking process
and the functions $f_{q,d}$ with $d \in D$. The set of terms $T_q$ whose
elements $t$ maximize the expression $\simil(f_{q,d},f_{t,d})$ is computed.
It contains the terms for all documents in $D$ that have a similar distribution
as the query, i.e.\ terms positioned near the query in the top ranked documents.
This set $T_q$ is used to expand $q$.

\begin{figure}[t]
\begin{tabular}{cc}
\includegraphics[width=0.44\linewidth]{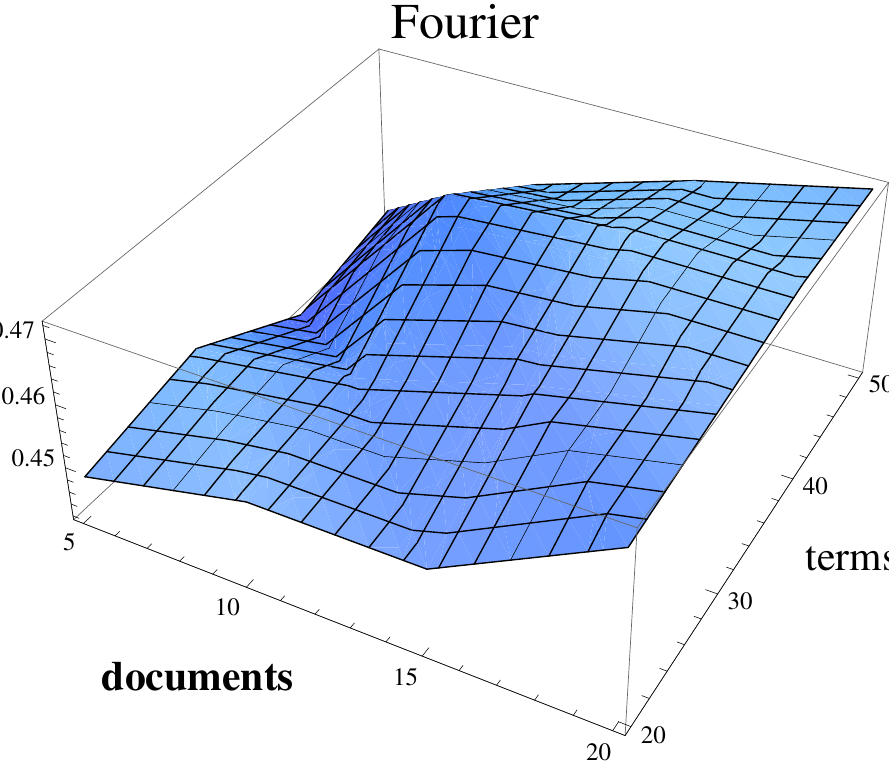} &
\includegraphics[width=0.44\linewidth]{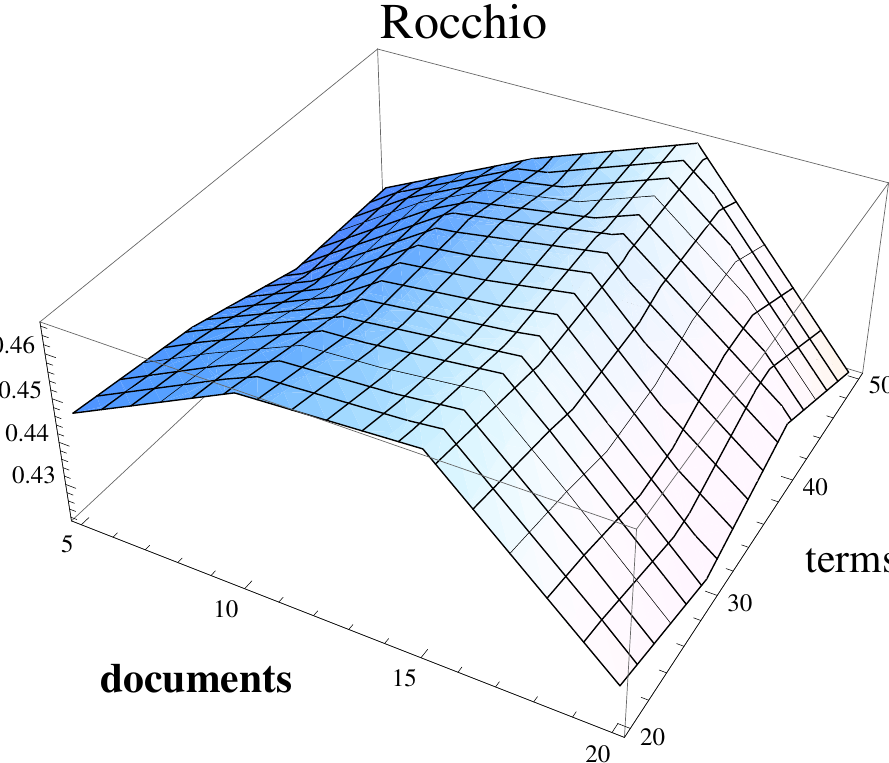} \\
\includegraphics[width=0.44\linewidth]{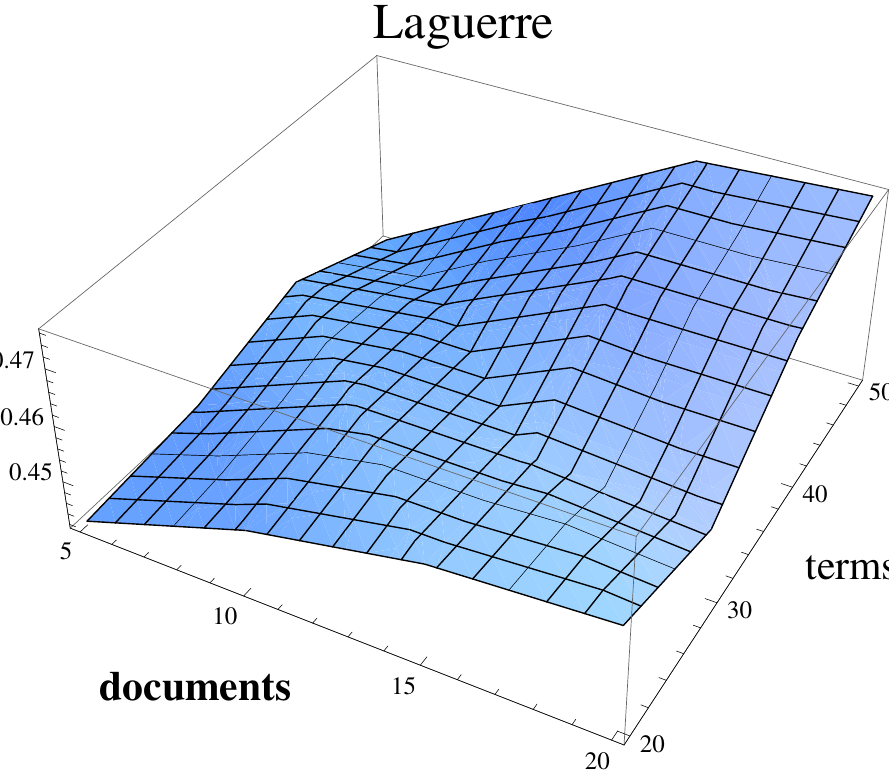} &
\includegraphics[width=0.44\linewidth]{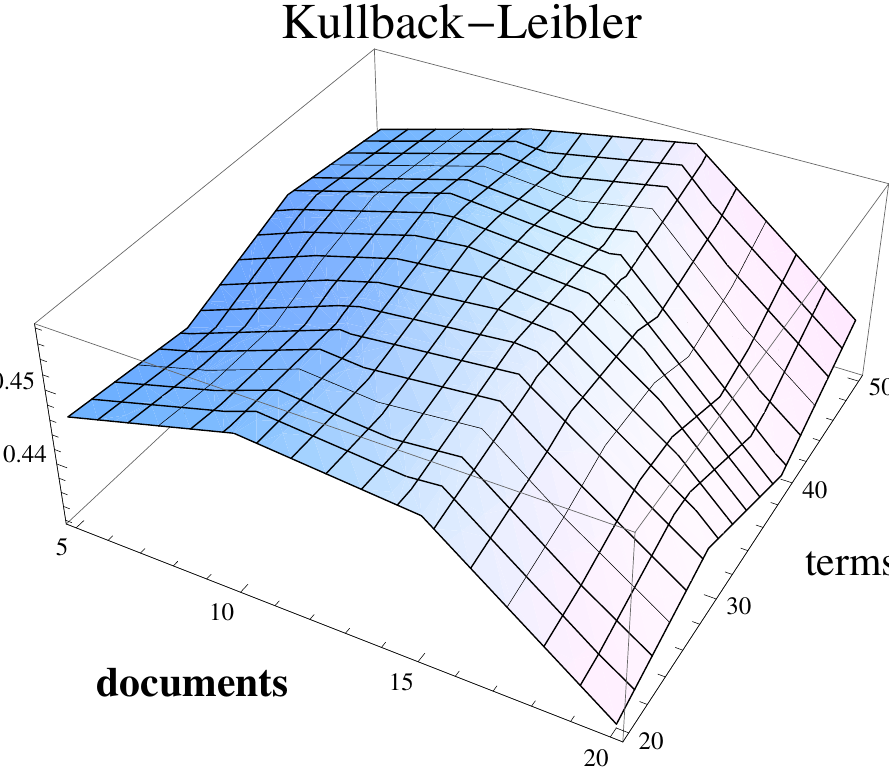}
\end{tabular}
\caption{Precision at 10 documents for term positional models and two other models,
using different query expansions configurations. The axes labeled documents and terms
correspond to $|D|$ and $|T_q|$, respectively.}
\label{configuring-expansion}
\end{figure}

As depicted in Figure \ref{configuring-expansion}, experiments executed on
the TREC-8 collection demonstrate that query expansion based on the proposed
orthogonal functions (Fourier and Laguerre) outperform state-of-the-art query
expansion models, such as Rocchio and Kullback-Leibler \cite{Ounis06}.
The term position models (left) differ from the other models (right) because the former
tend to increase the retrieval performance by increasing the number of
expansion documents and expansion terms, while for the other models, the performance
drops beyond roughly the $15^{th}$ expansion document.

Figure \ref{word-sphere-and-global-evaluation} (left) shows a fixed query expansion configuration
in which the other models show their best performance. Nevertheless, the term
distribution models perform better. Any increase in the number of expansion
documents or expansion terms makes the superiority of the term distribution
models even clearer.

\subsection{Cluster Analysis of Terms in Documents}
\label{clusters}

Given a document, one may ask whether there are groups (clusters) of terms whose
elements all have similar distributions. One may then infer that all terms
inside a cluster describe related concepts \cite{Attar77}. In this
section, some properties of the proposed method will be explained that
may be useful for the analysis of term clusters.

Consider a document of length $L$. Since at every position within the document
a particular term may either be present or not, there are in total $N = 2^L$
possible term distributions. Each of these distributions is mapped to a point
in an $(n + 1)$-dimensional Hilbert space. If the norm difference (\ref{eq12})
is used as the similarity criterion, then clusters of similar term distributions
are just Euclidean point clusters in the Hilbert space.

We will now investigate the geometrical structure of the set of all possible
term distributions. Let us first calculate the center
$\bar{f}(x) = (1 / N) \sum_{\nu = 1}^N f^{(\nu)}(x)$ of all term distributions
(here $f^{(\nu)}(x)$, $\nu=1,\ldots,N$, is an enumeration of distributions of the form (\ref{eq5})). At any position $x$, half of all $N$ distributions have a term
present ($f^{(\nu)}(x) = 1$) and the other half does not ($f^{(\nu)}(x) = 0$), so
that $\bar{f}(x) = 1 / 2 = \const$ for all $x \in [0, L]$. This average
distribution is mapped to a non-truncated, in general infinite-dimensional
coefficient vector $\vec{\bar{c}}$, whose length $|\vec{\bar{c}}|$ is given by
the norm $\|\bar{f}\| = [\int_0^L \md x / 4]^{1 / 2} = \sqrt{L} / 2$. The
squared distance between the center point and the coefficient vector
$\vec{c}^{(\nu)}$ of a distribution $f^{(\nu)}$ is
$|\vec{\bar{c}} - \vec{c}^{(\nu)}|^2  =  \|\bar{f} - f^{(\nu)}\|^2
 = \int_0^L (1 / 2 - f^{(\nu)}(x))^2 \md x$. Since $f^{(\nu)}(x)$ is either 0 or
1, it follows that $(1 / 2 - f^{(\nu)}(x))^2 = 1 / 4 = \const$ for all
$x \in [0, L]$, giving $|\vec{c}^{(\nu)} - \vec{\bar{c}}| = \sqrt{L} / 2$ for
all $\nu$. This means that the non-truncated coefficient vectors of all term
distributions lie on the surface of a sphere with radius $\sqrt{L} / 2$ whose
center is at $\vec{\bar{c}}$. Because $|\vec{\bar{c}}| = \sqrt{L} / 2$, this
sphere touches the origin of the Hilbert space.

\begin{figure}[t]
\begin{center}$
\begin{array}{cc}
\includegraphics[width=0.60\linewidth]{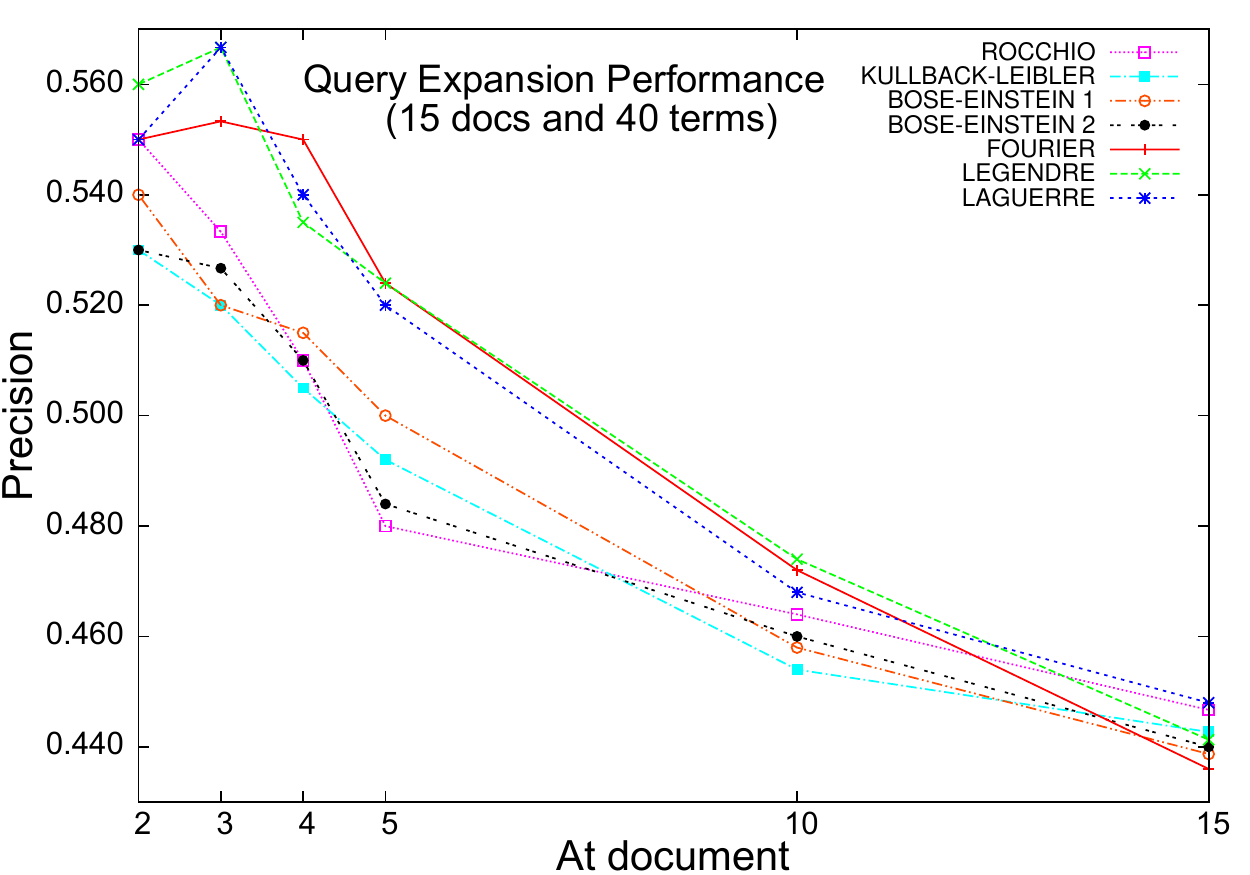} &
\includegraphics[width=0.46\linewidth]{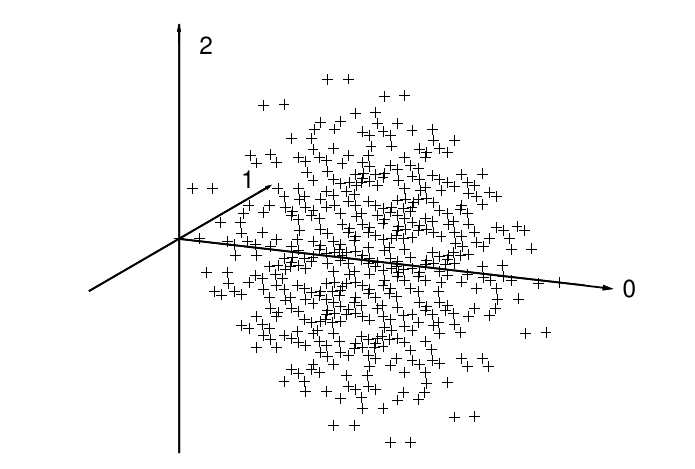}
\end{array}$
\end{center}
\caption{\textit{Left}: Query Expansion performance for the term distribution models (Fourier,Legendre and Laguerre) and the other models, using a configuration of 15 expanded documents and 40 expanded terms. \textit{Right}: Three dimensional sphere of all 512 possible term distributions in a document of length $L=9$ for the expansion in terms of Legendre polynomials.}
\label{word-sphere-and-global-evaluation}
\end{figure}

Bessel's inequality (\ref{eq4}) leads to $|\vec{c}^{(\nu)}_n - \vec{\bar{c}}_n| \leq \sqrt{L} / 2$ for all $\nu$ for the coefficient vectors truncated to order $n$.
Thus, the truncated vectors all lie \textit{within\/} a sphere of radius
\beq
R_0 = \sqrt{L} / 2
\eeq
in the $(n + 1)$-dimensional Hilbert space. The center of this sphere is at
$\vec{\bar{c}}_n$. If---as in the Fourier and Legendre cases---one of the
expansion functions, say $\varphi_0(x)$, is constant, the vector $\vec{\bar{c}}$
describing itself a constant function has only a non-vanishing zero component:
$\vec{\bar{c}} = \vec{\bar{c}}_n = (\sqrt{L} / 2, 0, 0, \ldots)$.
Fig.~\ref{word-sphere-and-global-evaluation} (right) shows this term sphere in $n + 1 = 3$ dimensions for a
document of length $L = 9$ and the expansion in terms of Legendre polynomials.

The fact that all possible truncated coefficient vectors $\vec{c}^{(\nu)}_n$ lie
within a sphere whose radius and center are known is very useful for clustering
analysis. First of all, it shows where in the Hilbert space to look for
clusters. Secondly, assume one has found a cluster $K = \{\vec{k}_1, \ldots, \vec{k}_q\}$ of term distributions by some clustering algorithm (for an $n$th order truncation).
The volume of this cluster can be estimated by calculating the standard
deviation $R_K = [(1 / q) \sum_{i = 1}^q (\vec{k}_i - \vec{\bar{k}})^2]^{1 / 2}
= [(1 / (2 q^2)) \sum_{i, j = 1}^q (\vec{k}_i - \vec{k}_j)^2]^{1 / 2}$ (here
$\vec{\bar{k}}$ is the center of the cluster) and approximating the cluster by
a sphere of radius $R_K$. Since the volume of a sphere of radius $R_K$ in $n + 1$
dimensions is proportional to $R_K^{n + 1}$, the cluster occupies approximately a
part $\xi = (R_K / R_0)^{n + 1} = (2 R_K / \sqrt{L})^{n + 1}$ of the
theoretically available space. A cluster would then be considered as significant
only if $\xi \ll 1$.
An analysis of this kind may be useful to generate an ontology of terms based on
individual documents.

It has been conjectured that the use of quantum mechanical methods, in
particular infinite-dimensional Hilbert spaces and projection operators, may be
advantageous in IR \cite{Rijsbergen04}. The approach presented here
goes into this direction, because constructing appropriate sets of orthogonal
functions is a standard technique in quantum mechanics. Still, we emphasize
that our approach is essentially classical, not quantum mechanical, since it
does not use any of the interpretational subtleties of quantum mechanics.

\section{Conclusions}
\label{conc}
In this paper, a new approach to improve document relevance evaluation 
using truncated Hilbert
space expansions has been presented. The proposed approach is based on an abstract representation of term
positions in a document collection which induces a measure of proximity between
terms (semantic interaction range) and permits their direct and simple
comparison.
Based on this abstract representation, it is possible to shift the complexity
of processing term-positional data to the indexing phase, permitting the use
of term-positional information at query time without significantly
affecting the response time of the system.
Three applications for IR were discussed: (a) ranking optimization based
on a user-defined term distribution function, (b) query expansion based on
term-positional information, and (c) a cluster analysis approach for terms
within documents.

There are several areas of future work. For example, (a) quantifying the effect of the abstract term positions representation in the index size, (b) measuring the effectiveness of the proposed clustering approach, and (c) studying objective functions in documents having homogeneous structures (forms) are some of the topics that should be investigated.

\bibliographystyle{abbrv}
\bibliography{ecir2010}

\begin{thebibliography}{1}

\bibitem{Abramowitz}
M.~Abramowitz, I.~Stegun, M.~Danos, and J.~Rafelski.
\newblock {\em Pocketbook of Mathematical Functions}.
\newblock H. Deutsch, 1984.

\bibitem{Attar77}
R.~Attar and A.~S. Fraenkel.
\newblock Local feedback in full-text retrieval systems.
\newblock {\em Journal of the ACM}, 24(3):397--417, 1977.

\bibitem{Beigbeder05}
M.~Beigbeder and A.~Mercier.
\newblock An information retrieval model using the fuzzy proximity degree of
  term occurences.
\newblock In {\em SAC '05: Proceedings of the 2005 ACM Symposium on Applied
  Computing}, pages 1018--1022, New York, NY, USA, 2005. ACM.

\bibitem{Galeas09}
P.~Galeas, R.~Kretschmer, and B.~Freisleben.
\newblock Document relevance assessment via term distribution analysis using
  {F}ourier series expansion.
\newblock In {\em JCDL '09: Proceedings of the 2009 Joint International
  Conference on Digital Libraries}, pages 277--284, New York, NY, USA, 2009.
  ACM.

\bibitem{Ounis06}
I.~Ounis, G.~Amati, V.~Plachouras, B.~He, C.~Macdonald, and C.~Lioma.
\newblock Terrier: A high performance and scalable information retrieval
  platform.
\newblock In {\em Proceedings of ACM SIGIR'06 Workshop on Open Source
  Information Retrieval (OSIR 2006)}, 2006.

\bibitem{Park04}
L.~A. Park, K.~Ramamohanarao, and M.~Palaniswami.
\newblock Fourier domain scoring: A novel document ranking method.
\newblock {\em Transactions on Knowledge and Data Engineering}, 16(5):529--539,
  May 2004.

\bibitem{Tao07}
T.~Tao and C.~Zhai.
\newblock An exploration of proximity measures in information retrieval.
\newblock In {\em SIGIR '07: Proceedings of the 30th Annual International ACM
  SIGIR Conference on Research and Development in Information Retrieval}, pages
  295--302, New York, NY, USA, 2007. ACM.

\bibitem{Rijsbergen04}
C.~J. {van Rijsbergen}.
\newblock {\em The Geometry of Information Retrieval}.
\newblock Cambridge University Press, New York, NY, USA, 2004.

\bibitem{Yosida80}
K.~Yosida.
\newblock {\em Functional Analysis}.
\newblock Springer, 1980.

\end{thebibliography}

\end{document}